\documentclass{aa}
\usepackage{txfonts}
\usepackage{graphicx}
\usepackage{setspace}
\usepackage{natbib}
\usepackage{sidecap}
\bibpunct{(}{)}{;}{a}{}{,}
\usepackage{color}
\usepackage{subfigure}
\usepackage{floatflt}

\newcommand{\rul}{\rule[-2.50mm]{0mm}{7mm}}

\begin{document}

\title{VLBI phase--reference observations of the gravitational lens JVAS B0218+357}
\author{Rupal Mittal\inst{1} \and Richard Porcas\inst{1}  \and Olaf
  Wucknitz\inst{2,}\inst{3} \and Andy Biggs\inst{3,}\inst{4} \and Ian Browne\inst{5} }
\institute{ MPIfR, Auf dem H\"ugel 69, 53121 Bonn, Germany
\and Universit\"at Potsdam, Institut f\"ur Physik, Am Neuen Palais 10, D-14469
Potsdam, Germany  
\and JIVE, 7990 AA Dwingeloo, Netherlands
\and UK Astronomy Technology Centre, Royal Observatory, Edinburgh, U.K.
\and JBO, Macclesfield, Cheshire SK11 9DL, U.K.
}

\date{Received/Accepted}

\abstract{ We present the results of phase--referenced VLBA+Effelsberg observations at five
frequencies of the double--image gravitational lens JVAS B0218+357, made to establish the precise
registration of the A and B lensed image positions. The motivation behind these observations is to
investigate the anomalous variation of the image flux--density ratio (A/B) with frequency -- this
ratio changes by almost a factor of two over a frequency range from 1.65~GHz to 15.35~GHz. We
investigate whether frequency dependent image positions, combined with a magnification gradient
across the image field, could give rise to the anomaly. Our observations confirm the variation of
image flux--density ratio with frequency. The results from our phase--reference astrometry, taken
together with the lens mass model of \citet{Wucknitz2004}, show that shifts of the image peaks and
centroids are too small to account for the observed frequency--dependent ratio.  }

\maketitle

\section{Introduction}

\begin{figure}
  \centering
  \includegraphics[width=0.4\textwidth]{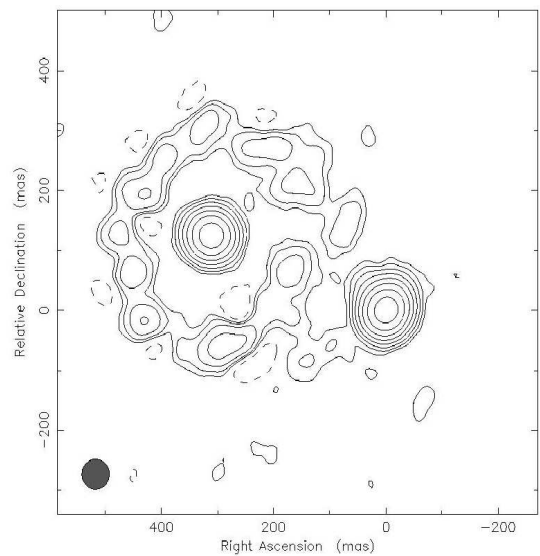}
  \caption{\small A combined EVN/MERLIN 1.7~GHz map of B0218+357, observed on
    ~$29^{\textrm{th}}$~May~1997 (Patnaik et al., unpublished). The resolution of the map is 50~mas
    and it shows an Einstein ring and the two compact images, A (right) and B (left). The contour
    levels are 6.5~mJy~$\textrm{beam}^{-1}~\times(-0.5,0.5,12,4,8,16,32,64)$ and the peak
    flux--density is 650~mJy~$\textrm{beam}^{-1}$.  }
\label{fig:EVN--MERLIN}
\end{figure}

\begin{table*}
  \caption{\small The image flux--density ratios as a function of frequency. S$_{\textrm{A}}$ and
    S$_{\textrm{B}}$ denote flux--densities of image A and B, respectively. Entries marked with an *
    indicate MERLIN and VLA measurements free from contamination by the Einstein ring. The error
    bars are included where available.}  \centering
  \begin{tabular}{|c|c|c|c|c|c|}
    \hline
    $\nu$ (GHz) & Interferometer & Resolution (mas$^2$) &  S$_{\textrm{A}}$ \rul (mJy) & 
    S$_{\textrm{B}}$ (mJy)  &  $\quad\textrm{S}_{\textrm{A}}/\textrm{S}_{\textrm{B}}$       \\  
   \hline

    1.7      & VLBI$^a$   &    5$\times$5    & 445$\pm$23  &  170$\pm$9  &  2.62$\pm$0.19              \\ 

    $5^*$    & MERLIN$^b$ &   50$\times$50   & 728$\pm$15  &  245$\pm$5  &  2.97$\pm$0.09              \\

    $4.84^*$ & VLA$^c$    &  400$\times$400  & 880$\pm$80  &  370$\pm$20 &  2.38$\pm$0.25              \\

    5        & EVN$^b$    &    5$\times$5    & 660         &  210        & 3.14                        \\

    5        & VLBI$^a$   &    1$\times$1    & 515$\pm$26  &  196$\pm$10 &  2.62$\pm$0.19              \\

    $5^*$    & MERLIN$^b$ &   50$\times$50   & 694$\pm$14  &  215$\pm$5  &  3.23$\pm$0.10              \\ 

    $8.4^*$  & VLA$^b$    &  200$\times$200  & 767         & 236         & 3.25                        \\   

    $8.4^*$  & VLA$^d$    &  200$\times$200  & 847         & 235          & $3.57\pm0.01$               \\

    $8.4^*$  & VLA$^c$    &  220$\times$220  &  807$\pm$40 &  271$\pm$15 &  2.98$\pm$0.22              \\

    $15^*$   & VLA$^d$    &  120$\times$120  & 935         & 570         & $3.73\pm0.01$               \\

    $15^*$   & VLA$^b$    &  120$\times$120  & 698         & 189         & 3.69                        \\

    $22.4^*$ & VLA$^c$    &   96$\times$76   & 833$\pm$160 &  253$\pm$50 &  3.29$\pm$0.90             \\

    \hline
  \end{tabular}
  \flushleft {\small $\qquad$ a $\;$ \citet{theory}\\ $\qquad$ b $\;$ \citet{1993Patnaik}\\ $\qquad$
      c $\;$ \citet{Dea1992}\\ $\qquad$ d $\;$ \citet{1999Biggs} (The flux--densities are averaged
      over the time--period of the observations, and the ratio is corrected for flux variability and
      the time delay; see text) \\ }
\end{table*}

The radio source JVAS B0218+357 was identified as a gravitationally lensed system by
\citet{1993Patnaik}. The source (Figure~\ref{fig:EVN--MERLIN}) consists of two compact images, A and
B, separated by $\sim$330~mas, the smallest angular separation amongst the known galaxy lenses, and
a faint Einstein ring of a similar diameter. Optical observations of the lens
\citep{1993Browne,Stickel1993,Dea1992} have revealed molecular absorption lines associated with the
lens galaxy at the redshift of 0.684. The morphology of the lens galaxy is spiral as suggested by
the large differential Faraday rotation ($\sim 900$~rad~$\textrm{m}^{-2}$) which is measured between
the images \citep{1993Patnaik} and detections of radio molecular absorption lines
\citep{Wiklind1995,Carilli1993} indicating that the lens--galaxy is rich in ionized gas, common to
late--type spirals. The small image separation, which is indicative of a low--mass galaxy acting as
a lens, lends support to this morphological categorization. The background source has properties of
a powerful radio source and a red optical continuum \citep{Stickel1993}, based on which it is
conjectured to be a blazar, and its redshift has been confirmed using optical spectroscopy \cite[z
$=$ 0.944,][]{2003Cohen,1993Browne}.  Further, it is variable in its radio emission and an accurate
value for the time delay between variations in the images has been measured by \citet{1999Biggs},
\mbox{(10.5 $\pm$ 0.4)~days} \citep[see also][]{2000Cohen}.

Previous VLBI observations at high frequencies \citep{2003Biggs,2001Kemball,1995Patnaik} show that
both images consist of a compact core component and a secondary compact component, offset by
ca. \mbox{1.5~mas} in the direction of a lower brightness jet of some 15~mas extent. At lower
frequencies \citep[e.g.][]{1996Porcas} the resolution is insufficient to separately resolve the two
compact components but the total extent of both images increases considerably with increasing
wavelength. In particular, image A, the stronger and larger, is extended in PA $-40^{\circ}$, the
expected direction of ``tangential stretching".

The wealth of data coming from numerous radio and optical observations of this system at various
frequencies and epochs provides constraints for lens models and, together with the time delay, give
the possibility to determine the Hubble constant, $H_0$. This has been accomplished by various
groups. \citet{1999Biggs} obtained a value of $H_0 = 69^{+13}_{-19}$~km~s$^{-1}$~Mpc$^{-1}$ using a
singular isothermal elliptical (SIE) mass distribution and \citet{Wucknitz2004} obtained a value of
$H_0 = 78^{+6}_{-6}$~km~s$^{-1}$~Mpc$^{-1}$ using a SIE potential (SIEP). The only problem in $H_0$
determination is related to the fact that the lens galaxy has weak optical emission, and is almost
completely masked by the emission from image~B (which is brighter than its counterpart in the
optical wavebands). Thus, the position of the lens galaxy is poorly known relative to the lensed
images and has been taken as a free parameter by the earlier authors in their (respective) lens
models.  A direct attempt to extract the optical centre of the lens--galaxy has been made by
\citet{York2005} who after subtracting the emission from the images, A and B, describe the lens
centre as a point about which the residuals are most symmetric. From their observations alone and
using a SIEP lens model, they derive $H_0 = 66^{+9}_{-9}$~km~s$^{-1}$~Mpc$^{-1}$ or $H_0 =
79^{+7}_{-7}$~km~s$^{-1}$~Mpc$^{-1}$ depending upon whether the spiral arms of the galaxy are masked
or not.

\begin{figure*}
  \centering
  \includegraphics[width=15cm,height=8cm]{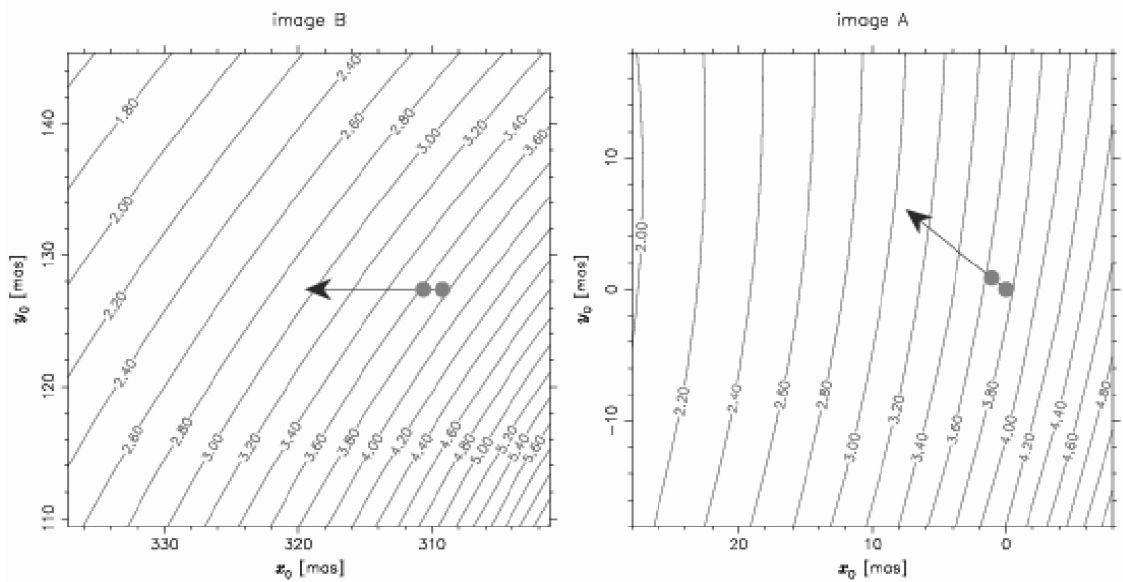}
  \caption{\small The curves indicate constant relative magnifications (A/B) for the best fitting
    lens model with the lens position at \mbox{$x_o = 260$~mas} and $y_o = 117.5$~mas
    \citep{2002phd}. The filled circles represent the sub--components, 1~(west) and 2~(east), and
    the arrows represent the expected direction of the shift in the centroid of the brightness
    distribution of the images at lower frequencies, i.e.\ towards sub--component 2 which forms the
    base of the radio--jet. }
  \label{fig:wucknitz}
\end{figure*} 

Even though the basic lensing characteristics of this system are well--understood and
well--reproduced by the current lens models, there are a few complications. One of them is the
steady and systematic decline in the ratio of radio image flux--density of the A and B images with
decreasing frequency. This is in direct contradiction to the expected achromatic behaviour of
gravitational lensing for point sources. Table~1 lists the observed ratios from previous
observations of B0218+357 at various frequencies and times, and using different interferometric
instruments. The ratio (A/B) varies from 3.7 to 2.6 over the frequency range 15.35~GHz to
1.65~GHz. At frequencies where the source is variable, the ratio measured at a single epoch will be
in error if the timescale for significant variations is comparable with the time delay. The
three--month VLA monitoring of B0218+357 at 15~GHz and 8.4~GHz by \citet{1999Biggs} has revealed
variations in the image flux--densities on a timescale of (70--100)~days. This, along with the
measured time--delay between the images, can induce variations in the ratio at the level of
(10--11)~\%. This is in good agreement with the 15~GHz ratios of the image flux--densities measured
at the same epoch which fall in the range from 3.45 to 3.96, with the mean value of 3.74 and an rms
of 0.02. At 8.4~GHz, the level of the observed variation in the ratio reduces to (7--8)~\% and the
rms reduces to 0.01. It is indeed expected that the variability amplitude of the background radio
source reduces with frequency as the contribution from non--variable radio components becomes
larger. Hence, although the time--variability in the background source emission along with the
time--delay between the lensed images results in a spread of the estimated image flux--density
ratios at a given frequency, it is clearly not the cause of the observed chromaticity in the
flux--density ratio.

We note that interferometric measurements may underestimate the individual image flux densities if
the beam size is much smaller than the extent of the image. If this occurs preferentially in one of
the images it will affect the flux--density ratio; since image A is larger than image B, this effect
might decrease the A/B flux--density ratio. Conversely, insufficient resolution may lead to
contamination of the B image by emission from the Einstein ring. We have included entries in Table~1
from short baseline instruments (VLA, MERLIN; marked with an `*') only at frequencies where the
resolution is sufficient to separate the ring emission from image B, to minimize this effect.

Although gravitational lens imaging of a point object is intrinsically achromatic, frequency
dependent variations in the image flux--density ratio of an extended source are possible, provided
its structure changes as a function of frequency and extends over regions of different relative
magnifications. It seems quite possible that this occurs in B0218+357 since both conditions
exist. Firstly, the small image separation leads to relative large changes in magnification across
the extent of the images.  For simple isothermal ellipsoid models of the lens potential, a shift of
15~mas in the position of a point--source image can produce a change in relative magnification from
4 to 2.5 (see Figure~\ref{fig:wucknitz}).

Secondly, the frequency--dependence of the B0218+357 image structures is strong, with previous VLBI
observations showing a marked increase of image sizes with decreasing frequency. Furthermore, it is
common for the radio spectra of one--sided AGN jets such as that seen in B0218+357 to steepen with
distance from the nucleus, providing a natural mechanism for producing a frequency--dependent
position.  The position of the radio peak at the base of the jet is also expected to move away from
the nucleus at lower frequencies -- the ``core--shift".

However, instrumental limitations prevent these effects from being seen easily.  One is the change
with frequency of the resolution available from VLBI observations. Another is the loss of absolute
position information when phase self--calibration is used to make VLBI maps, which prevents robust
registration of maps at different frequencies.  Frequency--dependent position shifts should, in
general, show up as a change with frequency of the separation between different images
\citep[e.g.][]{1995Porcas}. However, such differential shifts are hard to measure in some cases,
especially when the shift is in the same direction in both images.  Furthermore, any ``feature"
(e.g. the core) used to define the positions of the different images may, in fact, include different
fractions of the emission in the background source, since the images have different magnifications

In this paper we present the results of phase--referenced VLBI observations of the lensed source
B0218+357, made to establish an unambiguous registration of the structures of the radio images at
different frequencies. Although the expected core shifts are relatively small (mas), the positions
of the centroids of the image flux--density distributions may change with frequency by larger
amounts, and these are more representative of the position of the emission. It is therefore
important to quantify their changes with frequency when investigating the origin of the
frequency--dependent image flux--density ratios.

\section{Inverse phase--referencing}

VLBI observations suffer from the corrupting influence of the troposphere, ionosphere and
instrumental uncertainties on the interferometer visibility phase. It is well--known that the use of
phase self--calibration to eliminate these errors results in loss of the geometrical phases which
contain information regarding the source position relative to the antennas, resulting in the loss of
any information on the source position in the sky.

Alternatively, the technique of phase--referencing \citep{1989Alef} can be used, wherein
observations of the target and a reference source are alternated frequently, allowing mapping of the
target and determination of its position with respect to the reference.  If the reference source is
sufficiently compact that its position is achromatic, it can be considered as an astrometric
calibrator. Then, the phase--reference observations can be used to make a correct registration of
maps of the target source at different frequencies, to study any frequency--dependent structure.

Although this technique is most often used for mapping faint radio sources, the target in our study,
B0218+357, is sufficiently strong ($\sim$ 1 Jy) that we can use ``inverse phase--referencing".
Here, the target is used as the phase--reference for determining the corrupting phases, and
phase--reference maps of a point--like, achromatic astrometric reference source are made. One
advantage is that relatively faint sources can be used as astrometric references, permitting the
choice of sources closer to the target, which minimizes telescope drive times and reduces any
difference between the tropospheric and ionospheric phase corruptions of the target and
reference. Another advantage is that multiple astrometric reference sources can be used to guard
against the possibility that any single one may have chromatic structure.

\section{Observations}

VLBI observations were made at five frequencies, 15.35~GHz, 8.40~GHz, 4.96~GHz, 2.25~GHz and 1.65
GHz, using the NRAO Very Long Baseline Array (VLBA) and the Effelsberg radio telescope (Eb). Two
\mbox{14~h} source tracks were used ($13^{\textrm{th}}$ to $14^{\textrm{th}}$ and $14^{\textrm{th}}$
to $15^{\textrm{th}}$ January, 2002) and for each track (19~h to 03~h UT for Eb, 19~h to 09~h UT for
the VLBA) we switched between two receivers every 22~min. For the first track, observations were
alternated between 4.96~GHz and 2.25~GHz/8.40~GHz (S and X dual band) using \mbox{1~bit} sampling,
and for the second track between 1.65~GHz and 15.35~GHz (without Eb) using 2~bit sampling. The data
were recorded on four 8~MHz basebands at all the frequencies with the exception of eight basebands
at 4.96~GHz.  All observations were made in a single (LHC) circular polarization mode, except the
dual S/X observations which were made in RHC.

Apart from observing B0218+357, three compact reference sources were observed (see Table~2) along
with a fringe finder, B0234+285. Scans on B0218+357 of 1~min~20~s duration were interleaved by scans
of 2~min duration on each of the calibrators in succession. The position reference sources were
selected from the NRAO VLA Sky Survey (NVSS) catalogue. From an initial sample of candidate sources,
these three were found to exhibit flat spectra and the most point--like structure on the basis of
EVN and MERLIN 5~GHz observations \citep{2001Porcas}. They all lie within two degrees of B0218+357
and are stronger than 25 mJy at both 1.4~GHz and 5~GHz, with spectral indices flatter than
$-0.47$~(where the flux density, S~$\propto \nu^{\alpha}$; $\nu$ being the frequency and $\alpha$
the spectral index) .

The data were correlated at the VLBA correlator, with an output averaging time of 2.1~s and a
frequency resolution of 0.5~MHz, and further processed in AIPS, the Astronomical Image Processing
Software package provided by NRAO.

\begin{table}
  \caption{\small The position references, their separations from B0218+357, the correlated flux
    density on the short EVN baselines (S--EVN) and on the long EVN baselines (L--EVN) at 5~GHz. The
    ratio L/S is a measure of source--compactness.}  \centering
  \begin{tabular}{|c|c|c|c|c|}
    \hline
    Source name & Separation & S--EVN & L--EVN & L/S\\
    & (degrees) & (mJy) & (mJy) & \\
    \hline
    NVSS B0210+366 & 2.03 & 127 & 85 & 0.66\\
    NVSS B0215+364 & 1.03 & 140 & 86 & 0.61\\
    NVSS B0222+369 & 1.30 & 151 & 91 & 0.61\\
    \hline
  \end{tabular}
\end{table}

\begin{figure*}
  \centering
  \fbox{\includegraphics[width=17.5cm]{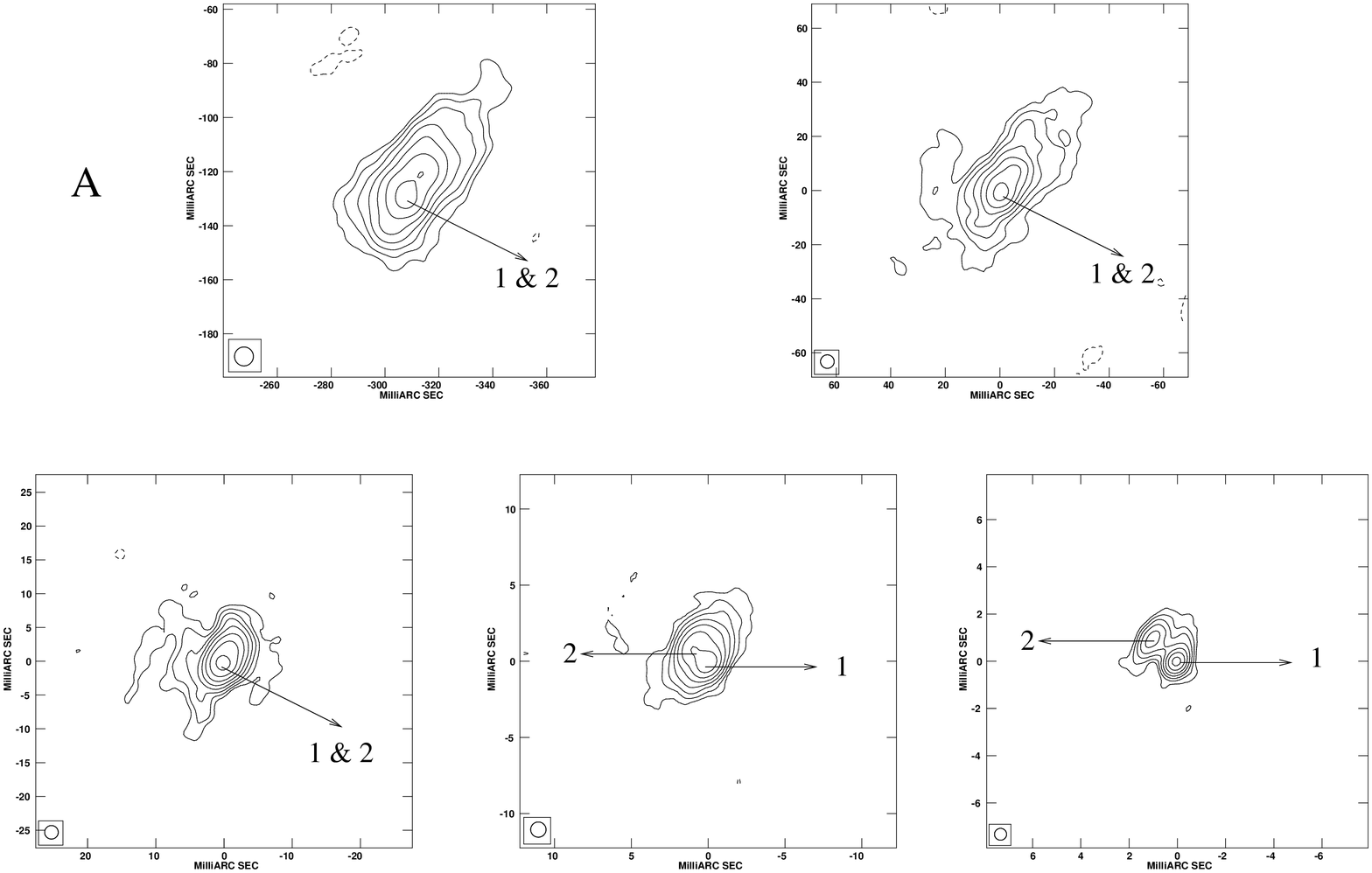}}
  \fbox{\includegraphics[width=17.5cm]{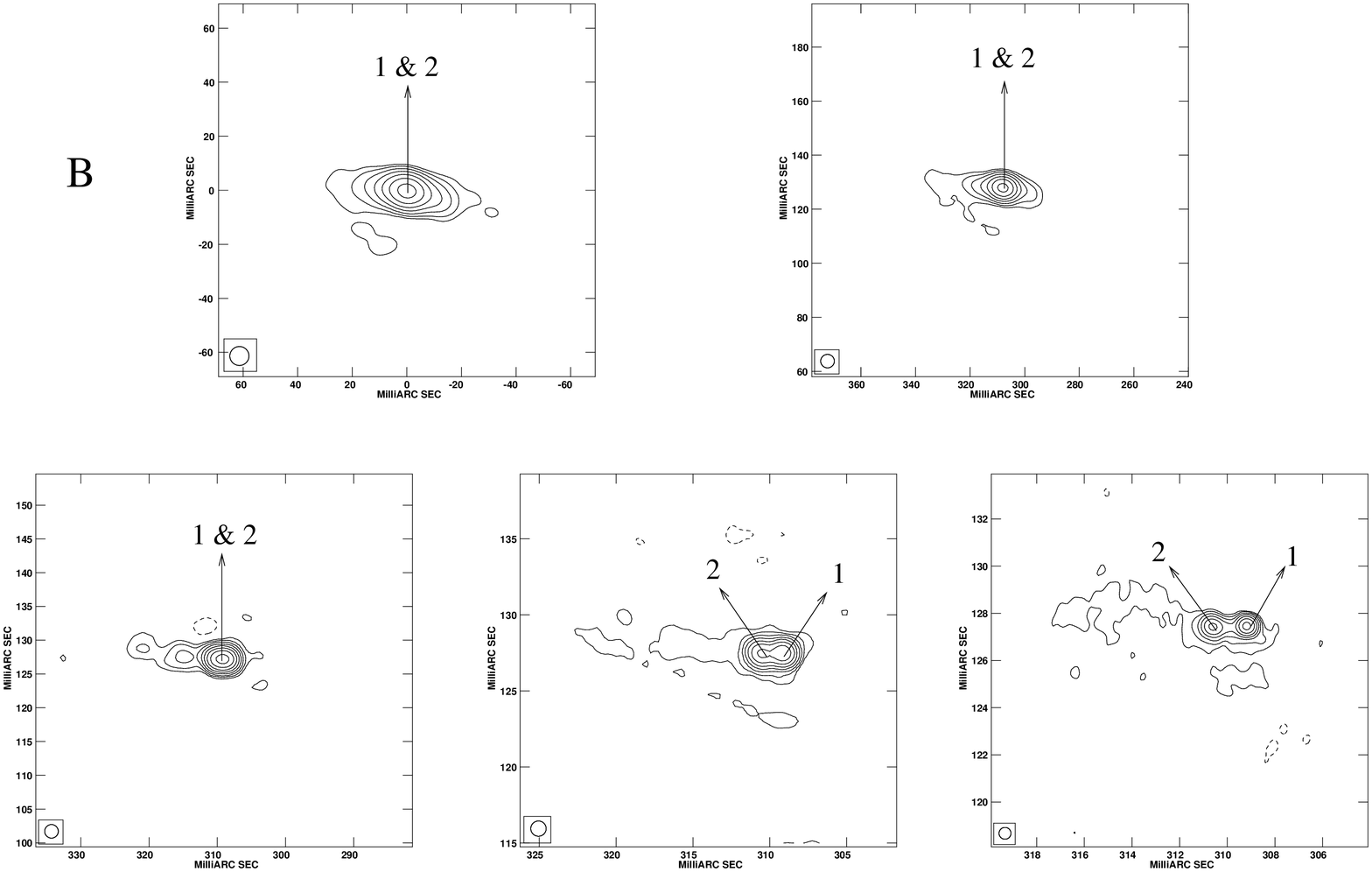}}
  \caption{\small The top and bottom panels show 1.65~GHz, 2.25~GHz, 4.96~GHz, 8.40~GHz and
  15.35~GHz maps of images A and B respectively, with frequencies increasing from left to right, top
  to bottom. The origin (0,0) is defined by the position of the peak intensity found in the initial
  run of fringe--fitting. This is the centre of A at all frequencies except for at 1.65~GHz at which
  the peak is in image B, entailing a reversal in the labeling of the axes. The residual noise in
  the maps is $\le$ 500 $\mu$Jy~$\textrm{beam}^{-1}$. The restoring beams used are (7, 5, 2, 1,
  0.5)~mas in the order of increasing frequency. The images clearly manifest all the earlier
  observed lensing--characteristics.  Image A is tangentially stretched at a PA $\sim$
  $-40\,^{\circ}$ and image B seems to be tangentially compressed. At 8.4~GHz and 15.35~GHz the
  images reveal further substructure, sub--components 1 and 2 with $\sim$ 1.4~mas separation,
  representing the core--jet morphology of the background source.  }
  \label{fig:images}
\end{figure*}

\section{Data analysis and maps}

\begin{figure}
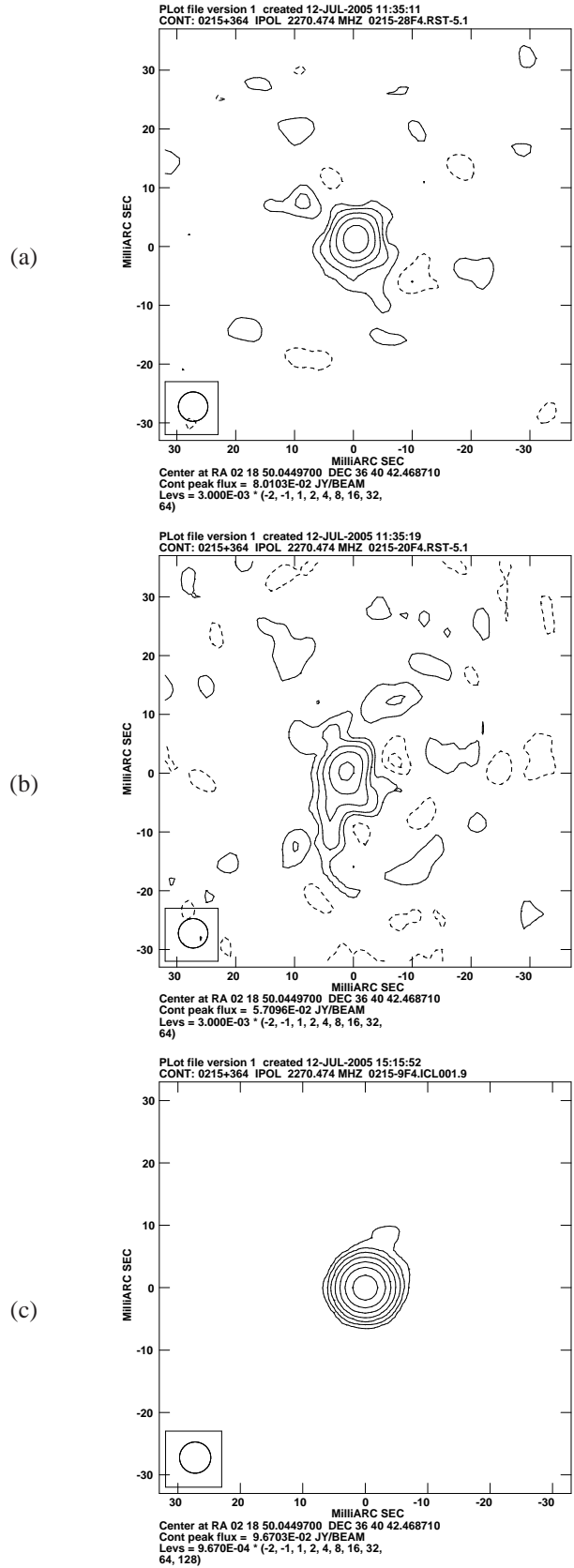

  \begin{minipage}{0.05\textwidth}
    \centering
    (a)
  \end{minipage}
  \begin{minipage}{0.45\textwidth}
    \centering
    \includegraphics[width=6.68cm]{0215_ionos.ps}
  \end{minipage}
  \begin{minipage}{0.05\textwidth}
    \centering
    (b)
  \end{minipage}
  \begin{minipage}{0.45\textwidth}
    \centering
    \includegraphics[width=6.68cm]{0215_no_ionos.ps}
  \end{minipage}
  \begin{minipage}{0.05\textwidth}
    \centering
    (c)
  \end{minipage}
  \begin{minipage}{0.45\textwidth}
    \centering
    \includegraphics[width=6.68cm]{0215-hybrid-9f4.ps}
  \end{minipage}
  \caption{\small Maps of 0215+364 at 2.3~GHz. a) Inverse phase--referenced map with ionospheric
    phase calibrations applied to the data. b) Without ionospheric phase calibrations. Both the maps
    are plotted with a 5~mas restoring beam and with identical contour levels at multiples of 3~mJy.
    c) Hybrid map obtained by means of self--calibration.}
  \label{fig:0215-ion}
\end{figure}

\begin{table*} 
  \caption{\small The image flux--density ratios from this data set. $b$ denotes the resolution and
    $\sigma$ is the root--mean--square noise in the maps.}  
  \centering
  \begin{tabular}{|c|c c|c|c|c|}
    \hline
    $\nu$ (GHz) & b (mas) & $\sigma$~(mJy/beam) & Flux Density:$\quad\textrm{S}^A$ (mJy) & Flux
    Density:$\quad\textrm{S}^B$ (mJy) & $\quad\textrm{S}^A/\textrm{S}^B$\\
    \hline

    1.65  &  50    & 0.35  &   $504\pm{9}$  &  $249\pm{5}$   &  $2.02\pm{0.05}$\\
    
    2.25  &  50    & 0.48  &   $646\pm{16}$ &  $243\pm{7}$   &  $2.67\pm{0.10}$\\
    
    4.96  &  30    & 0.25  &   $650\pm{9}$  &  $217\pm{3}$   &  $3.00\pm{0.06}$\\
    
    8.40  &  10    & 0.50  &   $681\pm{8}$  &  $206\pm{8}$   &  $3.31\pm{0.13}$\\
    
    15.35 &  5     & 1.00  &   $661\pm{5}$  &  $171\pm{6}$   &  $3.87\pm{0.14}$\\
    \hline
  \end{tabular}
\end{table*}
\begin{figure*}
  \subfigure[]{
    \label{}
    \begin{minipage}{0.5\textwidth}
      \centering
      \includegraphics[width=8.5cm]{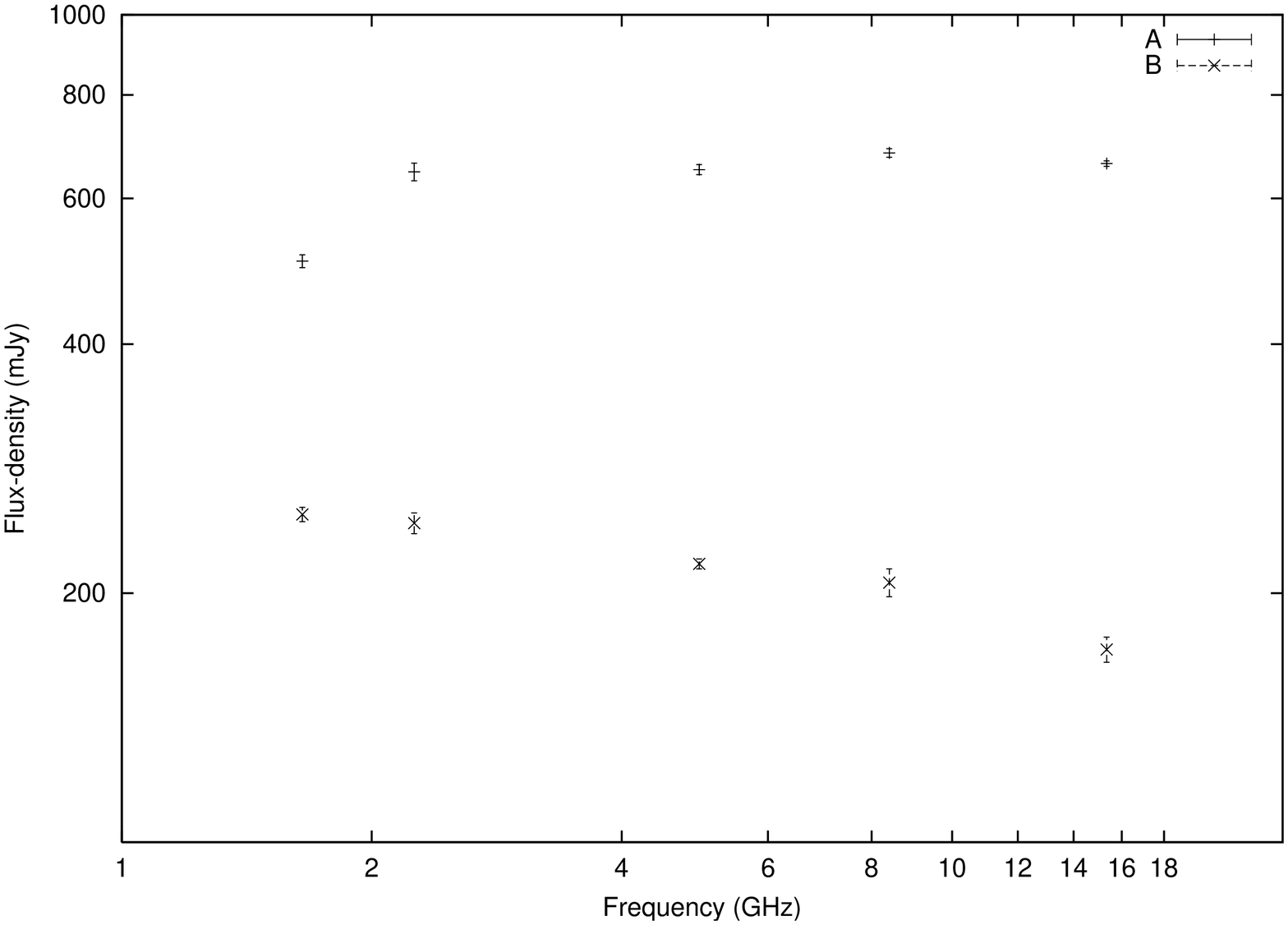}
    \end{minipage}
  }%
  \hfill
  \subfigure[]{
    \label{}    
    \begin{minipage}{0.5\textwidth}       
      \centering
      \includegraphics[width=8.5cm]{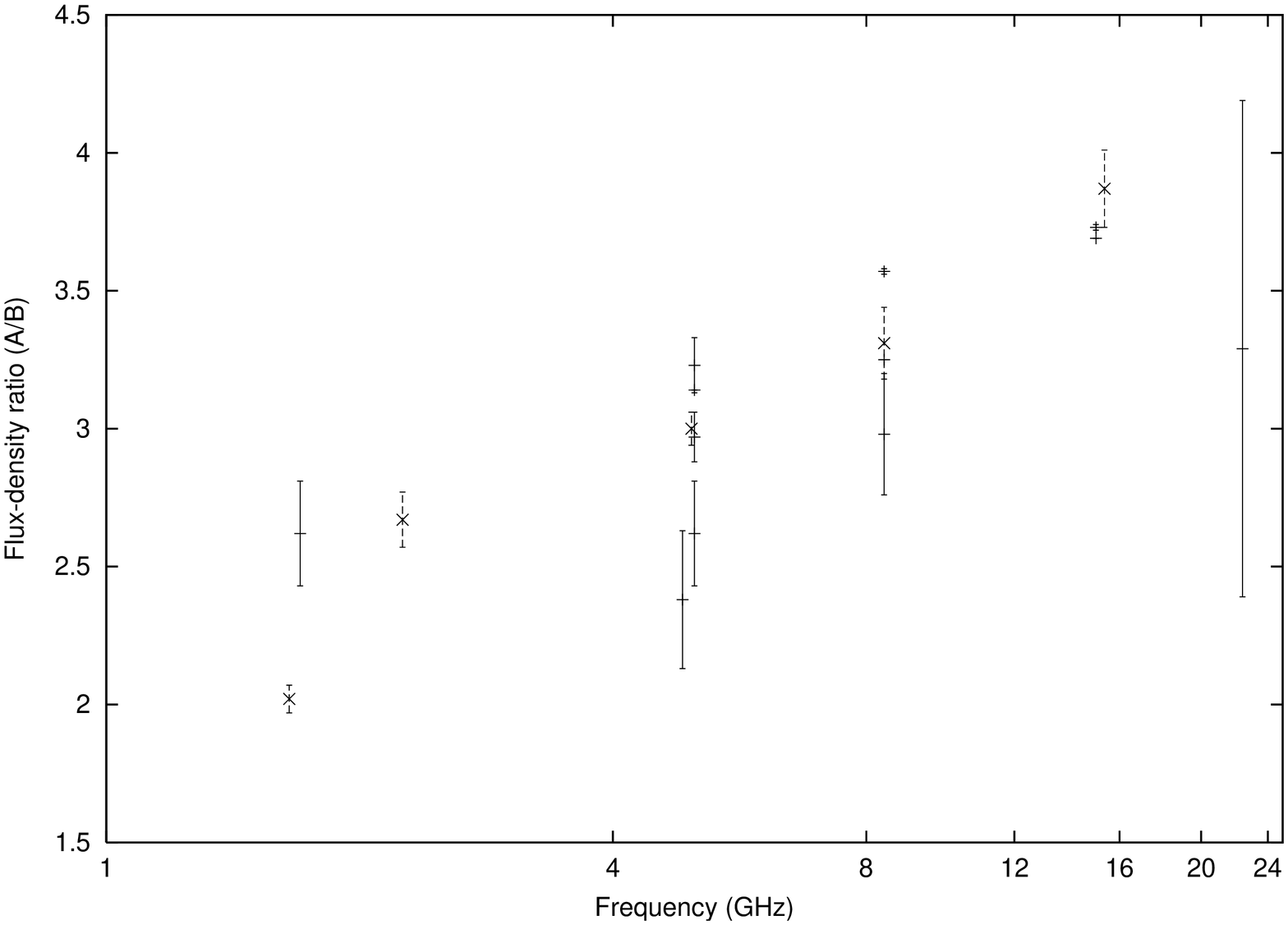}   
    \end{minipage}
  }%
  \caption{\small a) Image flux--densities obtained from these observations. b) Image flux--density
    ratio from these observations (with cross symbols and dashed error bars) and previous
    observations (with plus symbols and solid error bars, see Table~1).  }
  \label{fig:fluxandratios}
\end{figure*}

An essential step in the reduction of phase--reference data concerns the assumption of similar
tropospheric and ionospheric phase errors for the reference and the target source. For VLBI
observations the extra paths through the troposphere and ionosphere are very different for different
antennas, especially because the source is seen at different elevations. The mean phase errors due
to the troposphere, which scale as $\lambda^{-1}$, can be estimated approximately for each antenna,
and are taken into account in the model used for correlation. The mean phase errors due to the
ionosphere, which scale as $\lambda$, are highly unpredictable, however, and thereby prohibit the
use of any model at the time of correlation. Since the errors become pronounced at long wavelengths,
it is necessary to apply phase corrections from an ionospheric model after correlation. In applying
the ionospheric and tropospheric models, it is only the mean phase offset that is removed for every
antenna. The terms corresponding to small--scale temporal and spatial fluctuations still remain and
it is hoped they are the same for the target and the reference source. For these observations, we
used the AIPS task TECOR to apply an ionospheric model, produced by the Jet Propulsion Laboratory
(JPL), and which gives an estimate of the total electron content as a function of longitude,
latitude and time \citep[for more details see][]{1999Walker}.

In the further analysis standard reduction procedures were used, applying amplitude calibration
derived from telescope radiometry measurements and using fringe--fitting, self--calibration, CLEAN
deconvolution procedures and mapping. As mentioned earlier, B0218+357 is sufficiently strong that
self--calibration procedures could be applied at all frequencies. Hybrid maps were made by cleaning
two sub--fields simultaneously, one for image A and the other for B. After initial maps had been
obtained the fringe--fitting was repeated using these as an input model, and the mapping was
repeated. Full resolution hybrid maps of both A and B images at all five frequencies are shown in
Figure~\ref{fig:images}.

All three reference sources also proved to be strong enough to use phase self--calibration, and
hybrid maps were therefore also made for these. On inspection, two of these sources were deemed to
be unsuitable as position references. For B0210+366 the flux density drops to about 30 mJy at
1.6~GHz and consequently the position of the emission peak is not well defined. The maps of
B0222+369 reveal a jet like feature at 2.3~GHz and 1.6~GHz, casting doubt as to whether the position
of the peak is achromatic. Therefore for further analysis only B0215+364 was used as a position
reference; it appeared point--like in the maps at all frequencies. An amplitude self--calibration
procedure was used on this source to determine corrections to the {\it a priori} amplitude
calibration, and these were then applied to the B0218+357 data to improve the maps.

Phase--reference maps of B0215+364 were made at all frequencies by applying the self--calibration
solutions obtained from B0218+357 to the B0215+364 data. The dynamic range of these maps is poor and
deteriorates with increasing frequency from 30:1 to 2:1 with rms noise $\sim (1-1.5)$~mJy. To
investigate the efficacy of the ionospheric model we compared phase--referenced images made with and
without ionospheric phase corrections. One such comparison can be seen in Figure~\ref{fig:0215-ion}
which shows maps of B0215+364 at 2.25~GHz. The peak flux per unit beam is almost 1.5 times higher if
an ionospheric model is used [Figure~\ref{fig:0215-ion}(a)]. Moreover the peak positions differ by
almost half a milliarcsecond.  We conclude that even though the rms deviation in the total electron
content measured by different groups is on the order of 25~\%, the effect of applying any of the
models is in the direction that improves the results.

\section{Results}

\begin{figure}
  \centering
  \includegraphics[width=6cm]{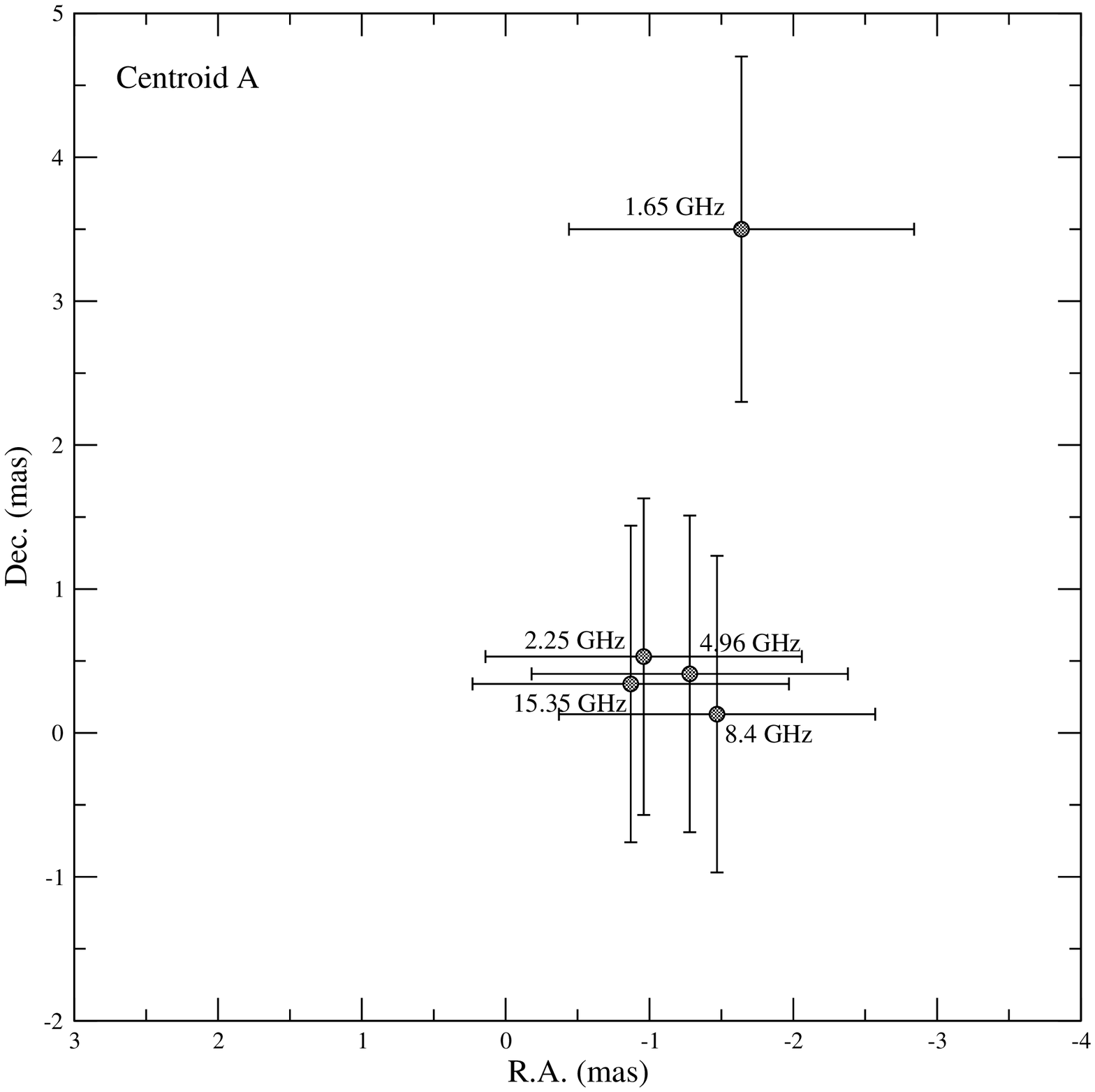}
  \hfill       
  \includegraphics[width=6cm]{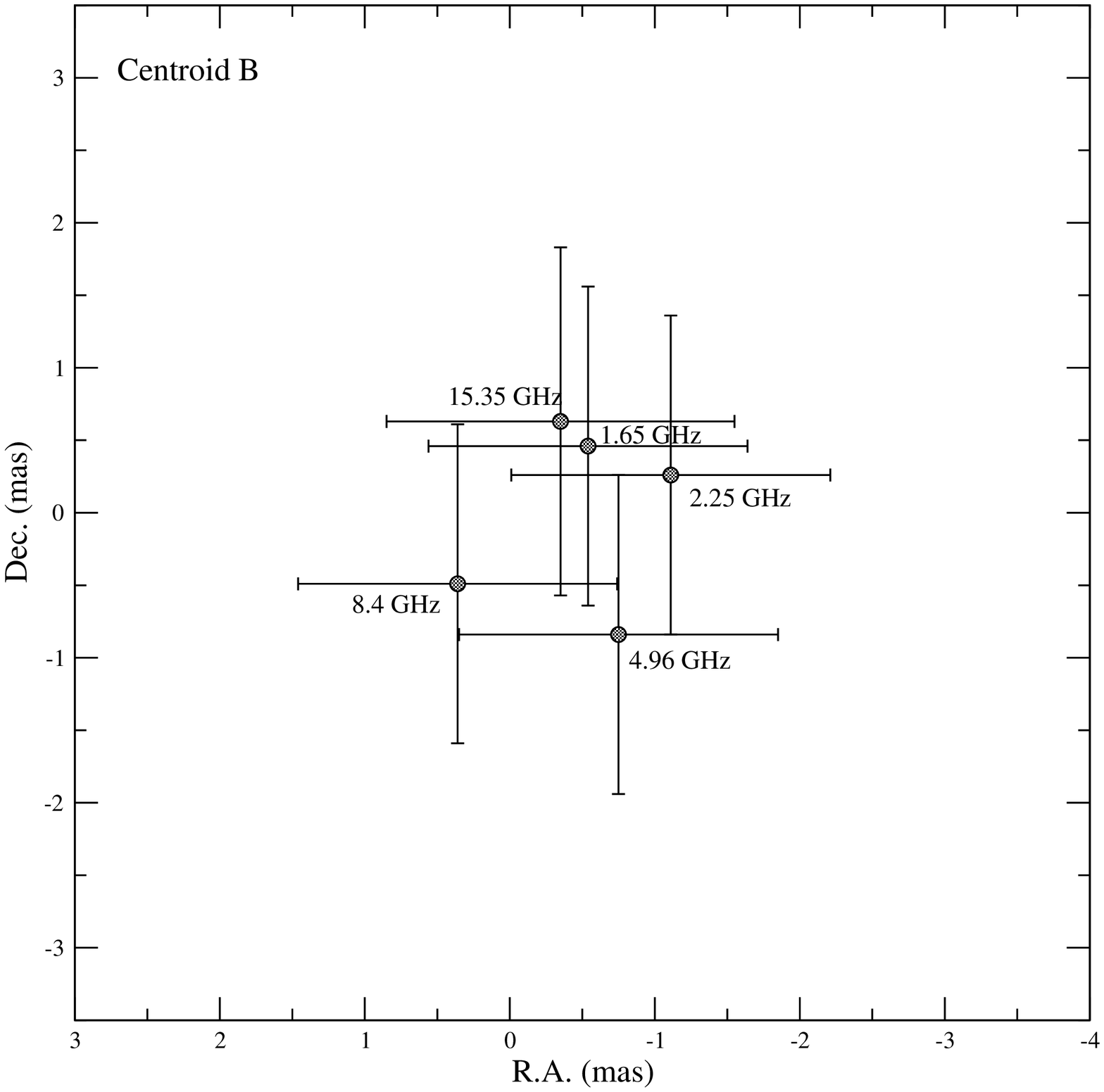}
  \caption{\small The change in the centroid position relative to the position--reference B0215+364,
    at five frequencies in image A (top) and image B (bottom). The error bars at 1.65~GHz, 2.25~GHz
    and 4.96~GHz are dominated by the position uncertainty in the B0218+357 maps due to the use of
    relatively large beams, and at 8.4~GHz and 15.35~GHz due to the position uncertainty of
    B0215+364 caused by the poor dynamic range of the phase--referenced maps at these frequencies. }
  \label{fig:phase-ref}
\end{figure}

The image flux--densities and flux--density ratios from these observations are presented in Table~3.
To guard against loss of flux density due to over--resolution, we made low--resolution maps by
discarding data from the longest baselines (using u,v resolution cut--offs). The flux densities from
these maps were estimated by putting a box around the images and integrating the flux density within
the box. The errors on the integrated flux--densities given in the table are derived from examining
the spread in estimates obtained by varying a number of parameters in the imaging process, and are
larger than the formal errors obtainable from a Gaussian fit. Shown in
Figure~\ref{fig:fluxandratios}(a) are the image spectra and it can be seen that image A
flux--density remains almost constant (to within $\pm$ 25 mJy) at the upper four frequencies but
drops suddenly by about 130~mJy (20 percent) at 1.65~GHz. In contrast, image B shows a gradual,
monotonic increase with wavelength with no sharp drop. We note that the variation of the image
flux--density ratio, as shown in Figure~\ref{fig:fluxandratios}(b), is similar in trend to that
found for previous observationsø, although the range is even larger ($\sim$~4 to $\sim$~2), with the
value at 1.65~GHz differing most from previous estimates.

The result of the phase--referenced observations is shown in Figure~\ref{fig:phase-ref}, which shows
the variation with frequency in the centroid position of the radio emission in the images with
respect to B0215+364. To determine an image centroid position, the CLEAN components were convolved
with a low--resolution restoring beam, and the peak in the resulting map was located using AIPS task
MAXFIT, which fits a quadratic function to the brightest area of the map to determine the position
offset with respect to the map centre.  The same procedure was used to locate the offset from the
map centre of the peak position of the calibrator, 0215+364, but using a full--resolution,
phase--referenced map.  The position of the B0218+357 image centroid with respect to 0215+364
(ignoring the constant position difference used for correlation) is then given by the difference of
these offsets (B0218+357 -- 0215+364). The same procedure was used for image B. The upper panel of
the figure indicates a relative shift of $\sim 5$~mas in the centroid of image A between 15.35~GHz
at 1.65~GHz, although the shift occurs only at 1.65~GHz. For image B, on the contrary, there is no
shift detected in the centroid of the brightness distribution. From the Singular Isothermal
Elliptical Potential (SIEP) model (Figure~\ref{fig:wucknitz}), we find that a shift of 5~mas may
cause the flux--density ratio to change as much as $\sim0.5$, and looking at Table~3, this
corresponds to $\sim$ 12.5~\% change in the observed flux--density ratio. However, this depends upon
the direction in which the shift has occurred.  From the phase--referenced results for image A, we
infer that this direction roughly coincides with the constant relative magnification contours (more
or less tangential with respect to the lens centre) and this shift corresponds to $\le$ 6~\% change
in the observed ratio. Therefore we draw the conclusion that the measured shift with frequency of
either image centroid positions is not sufficient to account for the flux--density ratio anomaly for
B0218+357.

\begin{figure}
  \centering
  \includegraphics[width=8cm]{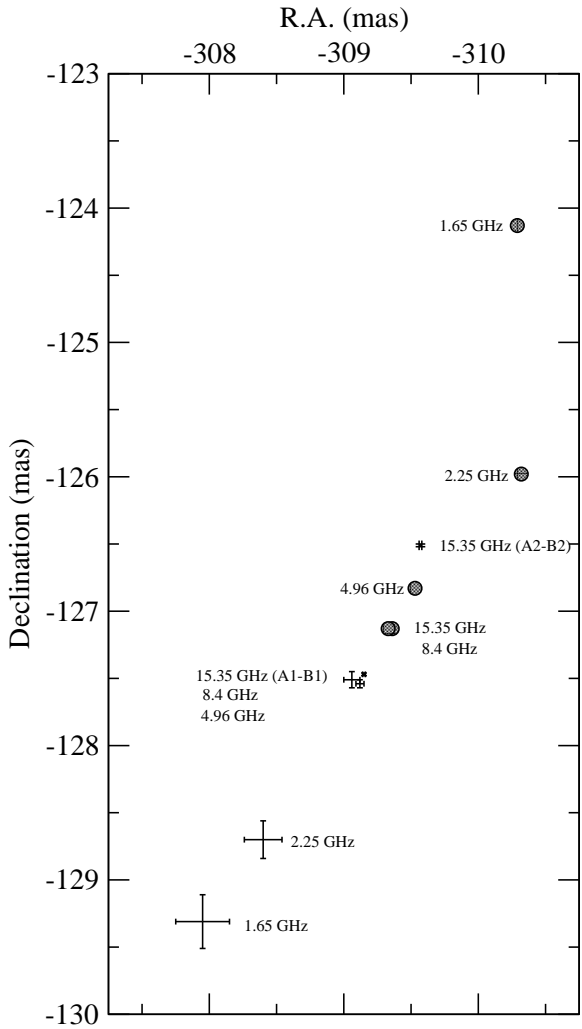}
  \caption{\small The A and B peak--to--peak separation (plus symbols) as a function of
    frequency. At \mbox{15.35~GHz}, separations for both the components A1$-$B1 and A2$-$B2 are
    marked. The A and B centroid--to--centroid separation is also shown (in circles) but the error
    bars are omitted for the sake of clarity.  }
  \label{fig:A-B-sep}
\end{figure}

An advantage of imaging gravitational lenses is that any frequency--dependent position in the source
can be seen as a frequency--dependence of the separation vector of the lensed images, without
requiring the use of an external reference source. As mentioned in Section 1, this method is
insensitive in cases where the position shift is along the same direction in both the images, but
the accuracy of the measurement of the A$-$B separation is much higher than for that between
B0218+357 and B0215+364. The peak--to--peak image separations at different frequencies, measured
with MAXFIT in full resolution maps, are shown in Figure~\ref{fig:A-B-sep}. Since the images at
15.35~GHz are resolved into sub--components 1 and 2, the separations of both A1$-$B1 and A2$-$B2 are
indicated in the plot. We see that the separation at 15.35~GHz (A1$-$B1), 8.4~GHz and 4.96~GHz
remains roughly the same, implying that the ``peak'' is centred around component 1 in both the
images. At 2.25~GHz and 1.65~GHz, the separation increases in declination and decreases in right
ascension, in a direction opposite to component 2. This is surprising as one might expect the
separation at lower frequencies to gradually coincide with the A2$-$B2 separation, reflecting
relatively more prominent emission from component 2 at the base of the jet. Shown in the same figure
as circles are the centroid--to--centroid image separations which are inconsitent with the
peak--to--peak separations, although the behaviour of the latter is in concert with the
image--structures seen at low frequencies (see below).

\begin{figure*}
  \subfigure[]{
    \label{fig:compt3A}
    \centering
    \begin{minipage}{0.5\textwidth}
      \includegraphics[width=7.5cm]{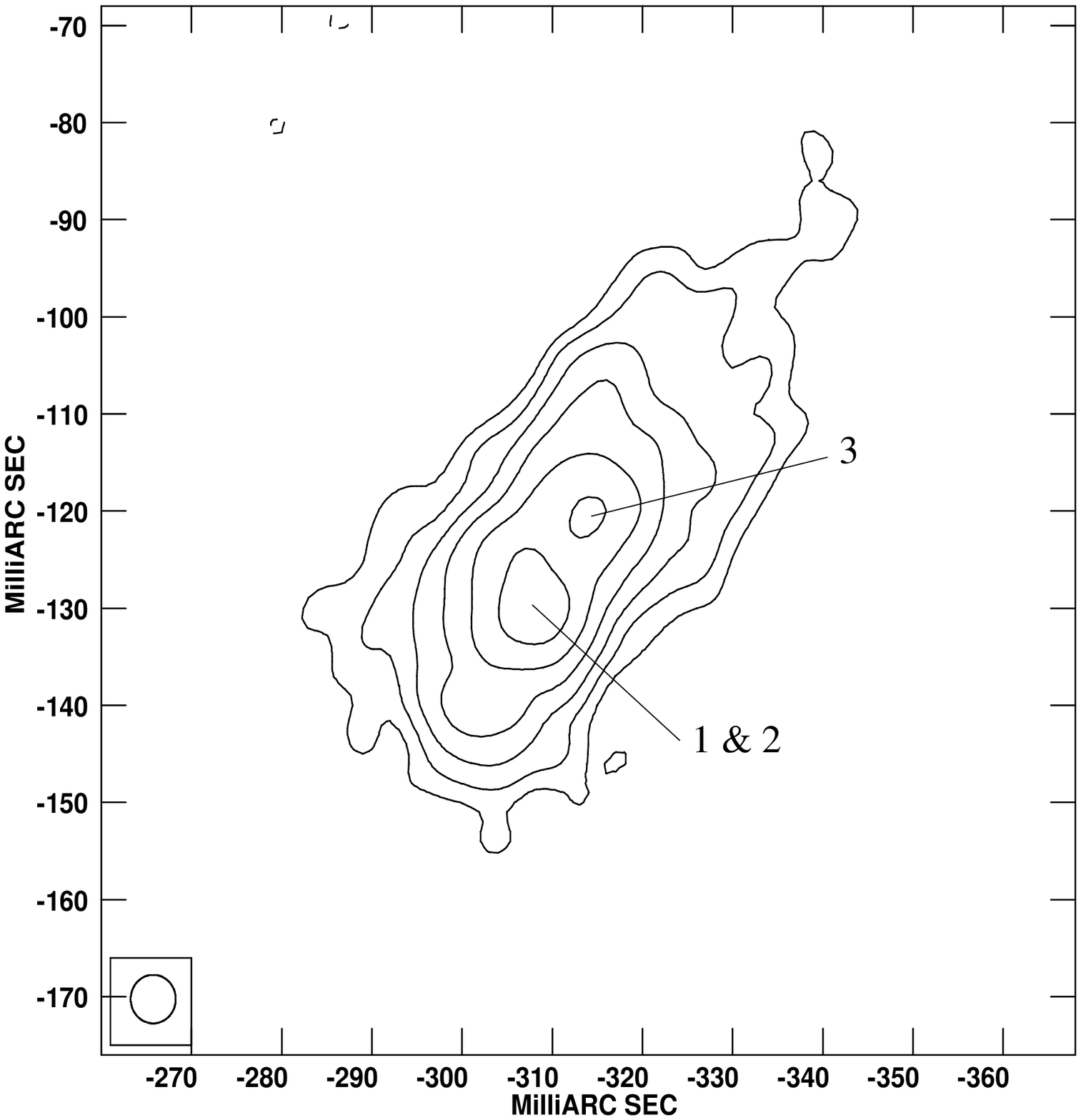}
    \end{minipage}
  }
  \hfill
  \subfigure[]{
    \label{fig:compt3B}
    \centering
    \begin{minipage}{0.5\textwidth}
      \includegraphics[width=7.5cm]{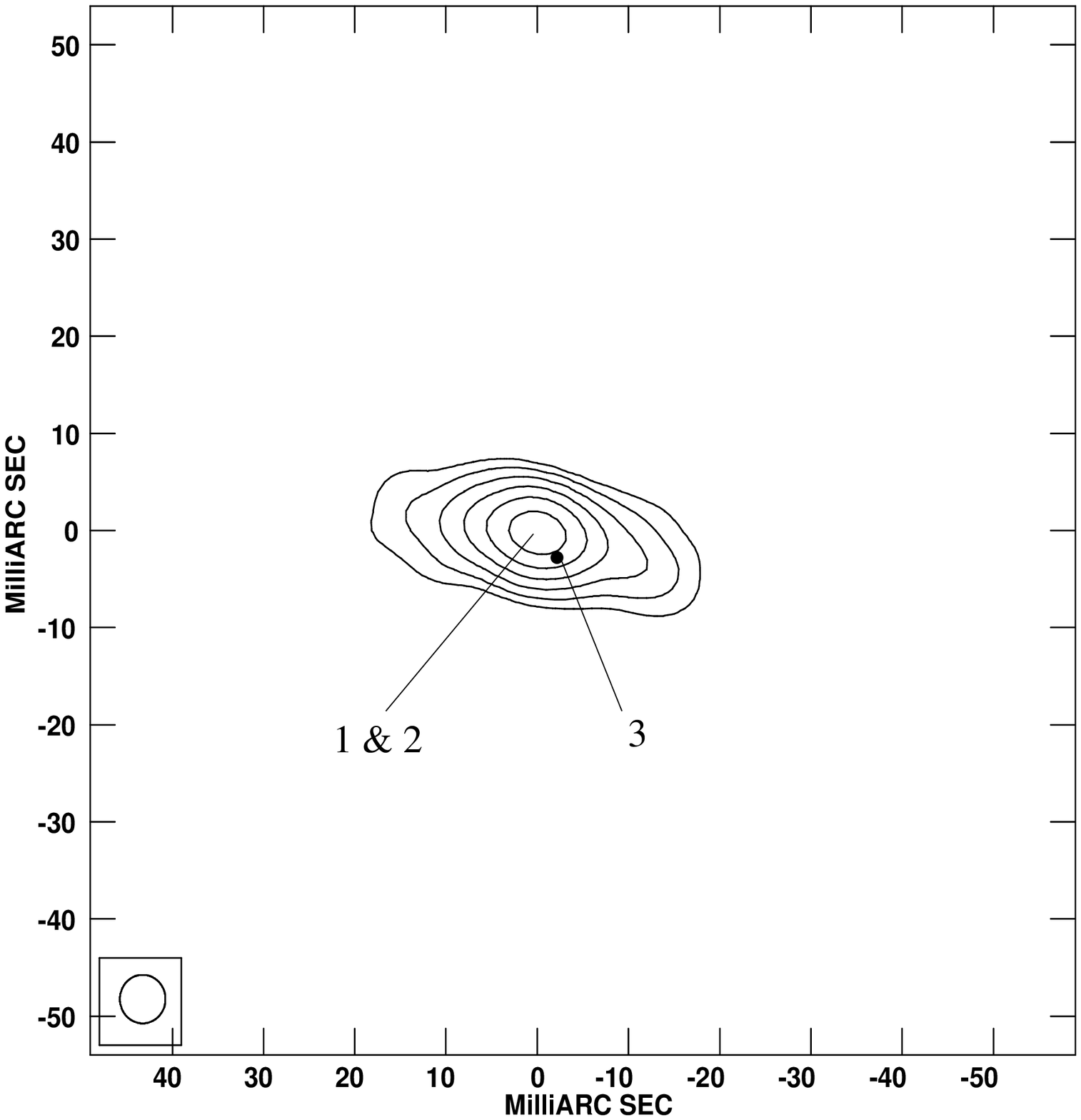}
    \end{minipage}
  }
  \subfigure[]{
    \label{fig:slice3A}
    \centering
    \begin{minipage}{0.5\textwidth}
      \includegraphics[width=7.5cm]{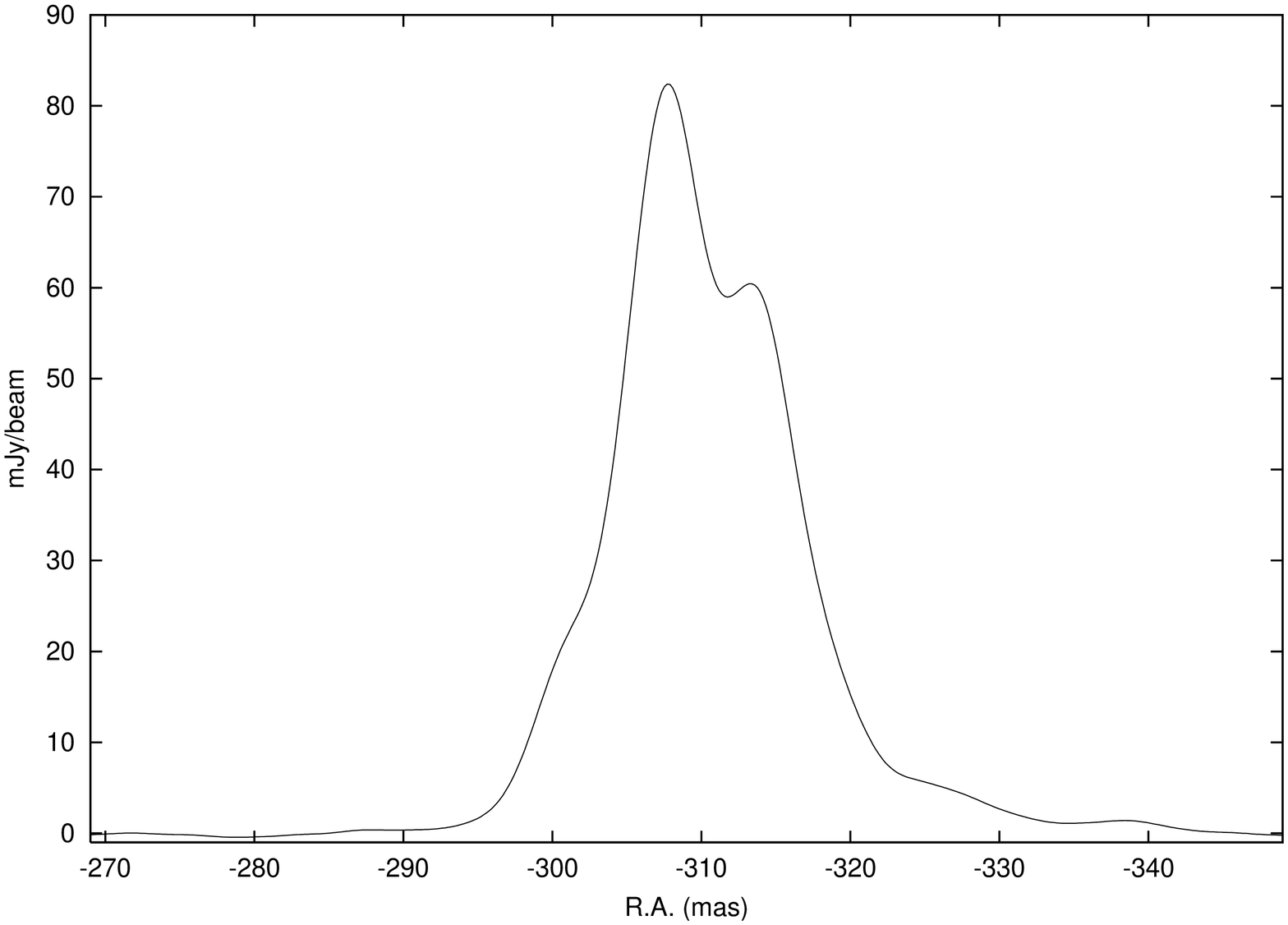} 
    \end{minipage}
  }
  \subfigure[]{
    \label{fig:slice3B}
    \centering
    \begin{minipage}{0.5\textwidth}
      \includegraphics[width=7.5cm]{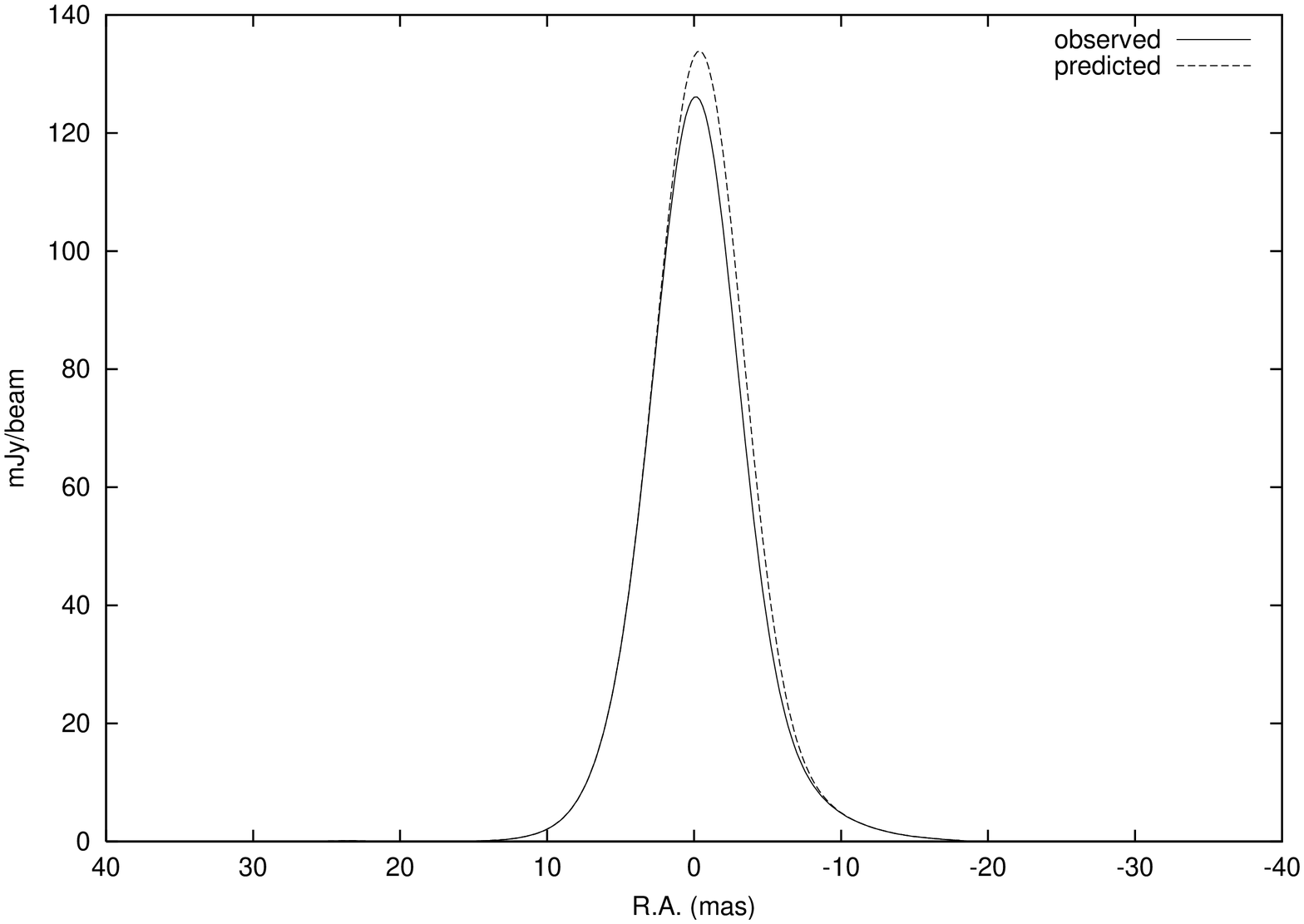}
    \end{minipage}
  }
  \caption{\small a) Second maximum (component 3) detected at 1.65~GHz in image A at a separation of
    $\sim10$~mas from component 1~\&~2 (The components 1 and 2 are not resolvable at these
    frequencies, hence what is marked is their superposition). b) The predicted position of
    component 3 in image B at 1.65~GHz, marked as a black dot about $\sim$ 3~mas from component
    1~\&~2. The resolution at this frequency is about 7~mas. c) A slice through the components
    1~\&~2 and 3 in image A. d) The predicted and observed slices through the components 1~\&~2 and
    the expected position of component 3 in image B.  }
  \label{fig:compt}
\end{figure*}

At 1.65~GHz we identify another component in image A, separated by $\sim$ 10~mas from component 1
and $\sim -33\,^{\circ}$ P.A., shown in Figure~\ref{fig:compt3A}. The origin of this
newly--identified feature (hereon component 3) is of great interest and we have investigated how
this can be part of the background source. On applying the lens model by \citet{Wucknitz2004}, we
deduce that in image B it should be $\sim$ 3~mas from component 1 as indicated in
Figure~\ref{fig:compt3B}.  Unfortunately the resolution at this frequency is not enough ($\sim$
7~mas) to resolve this separation. This can be seen by placing a component at the expected position
of component 3 in image B, of strength equal to the peak intensity of component 3 in image A
[Figure~\ref{fig:slice3A}] divided by the relative magnification ratio at this position. On
convolving this with the restoring beam, its effect can be assessed by comparing a slice made
through it and the components 1 \& 2 with the slice made through the same points in the observed
image. As shown in Figure~\ref{fig:slice3B}, including another component in image B analogous to
component 3 in image A has no observable effect other than changing the centroid of the brightness
distribution by distance less than the positional uncertainity. Therefore, we cannot distinguish
whether component 3 is a distinct feature in the background source imaged in A by the smooth macro
potential of the lens, or whether it is caused by some other mechanism. The shift in the centroid
position of image A at 1.6 GHz can be attributed to the existence of component 3.

\section{Discussion}

We have successfully used the technique of inverse phase--referencing to investigate the
frequency--dependence of the emission from the images of B0218+357; we believe this is the first
time in which a gravitational lens has been used as a phase--reference. We have established the
change in the centroid position of the image brightness distributions over five frequencies, and
also investigated the change in the positions of the emission peaks using inter--image
astrometry. The shift in the centroid in image A, which is significant ($\sim$ 5~mas) only at
1.65~GHz, is in a direction along which the relative magnification is predicted to be constant. In
image B no significant frequency dependent shift is detected in the position of the centroid. A
reasonable assumption is that the relative image magnification at the centroid positions derived
from the model gives a good measure of the expected image flux--density ratio. Thus we conclude that
the changing magnification gradient across the images is not the main cause of the anomalous change
of image flux--density ratio with frequency.

As can be clearly seen in Figure~\ref{fig:images}, at 1.65~GHz there is a large amount of low
brightness emission that extends out to $\sim \pm$ 30~mas. In comparison, at 15.35~GHz the emission
is dominated by the compact sub--components with a separation of $\sim$ 1.4~mas. At lower
frequencies the (larger) images extend over regions where lens models do, indeed, predict
significant changes in the relative magnification, and it may be insufficient to simply consider
magnifications at the centroid positions. This is because low frequency emission from different
regions of the background source is magnified by very different amounts. The resultant average
magnification is thus the integral of the (background source) intensity--weighted magnification over
the image area.

We plan to extend our study by applying the model from LensClean and evaluating the ratio of the
(background source) intensity--weighted magnifications of image A to B, explicitly taking the
observed extension of the background source into account. We also want to extend our choice of lens
model to one with a non--isothermal mass--radius profile. The motivation for this is the observation
by \citet{2003Biggs} related to the jet stretching factor, which seems to be larger in image B than
A, as seen when projected back to the source plane. But the deviation from non--isothermality is
expected to be small \citep{Wucknitz2004} since low--resolution VLA observations of the Einstein
ring are fitted quite well with isothermal models.

The identification of a distinct secondary maximum in image A at low frequencies (component 3), the
sharp downturn in the spectrum of image A at 1.6~GHz (not visible in image B) and the small shift in
the 1.6~GHz peak position in A are further results from these observations with no obvious
explanation in terms of the expected magnification gradients across the images. It may, therefore,
be necessary to consider more elaborate mechanisms, such as mass sub--structure (milli--lensing),
free--free absorption or refractive scattering in the lens galaxy, to account for all the observed
features in the B0218+357 images. Milli--lensing can produce significant frequency--dependent
changes in the flux--density of one (or both) of the images provided the source size is comparable
to the Einstein radius of the perturber and changes appreciably with frequency relative to it. The
source size changes by a factor 30 over the observed frequency range and its interaction with the
caustics of the perturber can bring about frequency--dependent changes in the total magnification
\citep{Dobler2005}. Non--gravitational effects such as free--free absorption or scattering will
cause flux--density perturbations that follow an inverse frequency power--law. The effects produced
by the above mechanisms on the image flux--density ratio in B0218+357 are a subject in their own
right and are beyond the scope of the present article \citep[][in preparation]{Mittal2006}.

\begin{acknowledgements}

We thank Martin Kilbinger and Alan Roy for useful suggestions and critiques, Peter Schneider for his
timely discussions, and the anonymous referee for a very beneficial report.  We would also like to
express our appreciation for the VLBA and Effelsberg operational staff, and the VLBA correlator
staff for the their help in obtaining the data. The National Radio Astronomy Observatory is a
facility of the National Science Foundation operated under cooperative agreement by Associated
Universities, Inc. The Effelsberg radio telescope is operated by the Max--Planck--Institut f\"ur
Radioastronomie (MPIfR). \,R.\,M. is supported by the International Max Planck Research School
(IMPRS) for Radio and Infrared astronomy at the Universities of Bonn and Cologne.

\end{acknowledgements}

\bibliographystyle{aa}
\bibliography{ref}

\begin{thebibliography}{23}
\expandafter\ifx\csname natexlab\endcsname\relax\def\natexlab#1{#1}\fi

\bibitem[{{Alef}(1989)}]{1989Alef}
{Alef}, W. 1989, in Very Long Baseline Interferometry, Techniques and
  Applications, NATO ASI Series C: Mathematical and Physical Sciences, ed.
  M.~{Felli} \& R.~E. {Spencer}, Vol. 283, 261

\bibitem[{{Biggs} {et~al.}(1999){Biggs}, {Browne}, {Helbig}, {Koopmans},
  {Wilkinson}, \& {Perley}}]{1999Biggs}
{Biggs}, A.~D., {Browne}, I.~W.~A., {Helbig}, P., {et~al.} 1999, \mnras, 304,
  349

\bibitem[{{Biggs} {et~al.}(2003){Biggs}, {Wucknitz}, {Porcas}, {Browne},
  {Jackson}, {Mao}, \& {Wilkinson}}]{2003Biggs}
{Biggs}, A.~D., {Wucknitz}, O., {Porcas}, R.~W., {et~al.} 2003, \mnras, 338,
  599

\bibitem[{{Browne} {et~al.}(1993){Browne}, {Patnaik}, {Walsh}, \&
  {Wilkinson}}]{1993Browne}
{Browne}, I.~W.~A., {Patnaik}, A.~R., {Walsh}, D., \& {Wilkinson}, P.~N. 1993,
  \mnras, 263, L32

\bibitem[{{Carilli} {et~al.}(1993){Carilli}, {Rupen}, \& {Yanny}}]{Carilli1993}
{Carilli}, C.~L., {Rupen}, M.~P., \& {Yanny}, B. 1993, \apjl, 412, L59

\bibitem[{{Cohen} {et~al.}(2000){Cohen}, {Hewitt}, {Moore}, \&
  {Haarsma}}]{2000Cohen}
{Cohen}, A.~S., {Hewitt}, J.~N., {Moore}, C.~B., \& {Haarsma}, D.~B. 2000,
  \apj, 545, 578

\bibitem[{{Cohen} {et~al.}(2003){Cohen}, {Lawrence}, \&
  {Blandford}}]{2003Cohen}
{Cohen}, J.~G., {Lawrence}, C.~R., \& {Blandford}, R.~D. 2003, \apj, 583, 67

\bibitem[{{Dobler} \& {Keeton}(2005)}]{Dobler2005}
{Dobler}, G. \& {Keeton}, C.~R. 2005, astro-ph/0502436 (submitted to \mnras)

\bibitem[{{Kemball} {et~al.}(2001){Kemball}, {Patnaik}, \&
  {Porcas}}]{2001Kemball}
{Kemball}, A.~J., {Patnaik}, A.~R., \& {Porcas}, R.~W. 2001, \apj, 562, 649

\bibitem[{{Mittal}(2006)}]{Mittal2006}
{Mittal}, R. 2006, Ph.D.~Thesis, Max--Planck Institute for Radioastronomy,
  Bonn, in preparation

\bibitem[{{O'Dea} {et~al.}(1992){O'Dea}, {Baum}, {Stanghellini}, {Dey}, {van
  Breugel}, {Deustua}, \& {Smith}}]{Dea1992}
{O'Dea}, C.~P., {Baum}, S.~A., {Stanghellini}, C., {et~al.} 1992, \aj, 104,
  1320

\bibitem[{{Patnaik} {et~al.}(1993){Patnaik}, {Browne}, {King}, {Muxlow},
  {Walsh}, \& {Wilkinson}}]{1993Patnaik}
{Patnaik}, A.~R., {Browne}, I.~W.~A., {King}, L.~J., {et~al.} 1993, \mnras,
  261, 435

\bibitem[{{Patnaik} \& {Porcas}(1999)}]{theory}
{Patnaik}, A.~R. \& {Porcas}, R.~W. 1999, in ASP Conf. Ser. 156: Highly
  Redshifted Radio Lines, ed. C.~L. {Carilli}, S.~J.~E. {Radford}, K.~M.
  {Menten}, \& G.~I. {Langston}, 247--251

\bibitem[{{Patnaik} {et~al.}(1995){Patnaik}, {Porcas}, \&
  {Browne}}]{1995Patnaik}
{Patnaik}, A.~R., {Porcas}, R.~W., \& {Browne}, I.~W.~A. 1995, \mnras, 274, L5

\bibitem[{{Porcas}(2001)}]{2001Porcas}
{Porcas}, R.~W. 2001, in 15th Working Meeting on European VLBI for Geodesy and
  Astrometry, 201--207

\bibitem[{{Porcas} \& {Patnaik}(1995)}]{1995Porcas}
{Porcas}, R.~W. \& {Patnaik}, A.~R. 1995, in 10th Working Meeting on European
  VLBI for Geodesy and Astrometry, 188--192

\bibitem[{{Porcas} \& {Patnaik}(1996)}]{1996Porcas}
{Porcas}, R.~W. \& {Patnaik}, A.~R. 1996, in IAU Symp. 173: Astrophysical
  Applications of Gravitational Lensing, ed. C.~S. {Kochanek} \& J.~N.
  {Hewitt}, 311--316

\bibitem[{{Stickel} \& {Kuhr}(1993)}]{Stickel1993}
{Stickel}, M. \& {Kuhr}, H. 1993, \aaps, 101, 521

\bibitem[{{Walker} \& {Chatterjee}(1999)}]{1999Walker}
{Walker}, C. \& {Chatterjee}, S. 1999, VLBA Scientific Memo, 23

\bibitem[{{Wiklind} \& {Combes}(1995)}]{Wiklind1995}
{Wiklind}, T. \& {Combes}, F. 1995, \aap, 299, 382

\bibitem[{{Wucknitz}(2002)}]{2002phd}
{Wucknitz}, O. 2002, Ph.D.~Thesis, University of Hamburg

\bibitem[{{Wucknitz} {et~al.}(2004){Wucknitz}, {Biggs}, \&
  {Browne}}]{Wucknitz2004}
{Wucknitz}, O., {Biggs}, A.~D., \& {Browne}, I.~W.~A. 2004, \mnras, 349, 14

\bibitem[{{York} {et~al.}(2005){York}, {Jackson}, {Browne}, {Wucknitz}, \&
  {Skelton}}]{York2005}
{York}, T., {Jackson}, N., {Browne}, I.~W.~A., {Wucknitz}, O., \& {Skelton},
  J.~E. 2005, \mnras, 357, 124

\end{thebibliography}

\end{document}